# A Hard-wall Potential Function for Transport Properties of Alkali Metals Vapor

*Mohammad Hadi Ghatee[*] and Fatemeh Niroomand-Hosseini*

(Department of Chemistry, Shiraz University, Shiraz, 71454, Iran)

*Author to whom correspondence should be addressed. Fax: +98 711 228 6008. E-mail: ghatee@susc.ac.ir.

**Abstract**

This study demonstrates that the transport properties of alkali metals are determined principally by the repulsive wall of the pair interaction potential function. The (hard-wall) Lennard-Jones(15-6) effective pair potential function is used to calculate transport collision integrals. Accordingly, reduced collision integrals of K, Rb, and Cs metal vapors are obtained from Chapman-Enskog solution of the Boltzman equation. The law of corresponding states based on the experimental-transport reduced collision integral is used to verify the validity of a LJ(15-6) *hybrid* potential in describing the transport properties. LJ(8.5-4) potential function and a simple thermodynamic argument with the input *PVT* data of liquid metals provide the required molecular potential parameters. Values of the predicted viscosity of monatomic alkali metals vapor are in agreement with typical experimental data with the average absolute deviation 2.97% for K in the range 700-1500 K, 1.69% for Rb, and 1.75% for Cs in the range 700-2000 K. In the same way, the values of predicted thermal conductivity are in agreement with experiment within 2.78%, 3.25%, and 3.63% for K, Rb, and Cs, respectively. The LJ(15-6) hybrid potential with a hard-wall repulsion character conclusively predicts best transport properties of the three alkali metals vapor.

**1. Introduction**

The transport properties of alkali metals vapor are very important as working fluid in industry[1,2]. The investigation of transport properties of alkali metal vapors is difficult because of high temperatures and low vapor pressure. On the other hand liquid alkali metals are corrosive and thus special technique is required for an accurate measurement. Under these conditions the experimental transport properties of alkali metal vapors in a wide range of temperature are scarce, and therefore, an accurate prediction is valuable for many industrial application[1,3,4].

The theoretical calculation of transport properties of a dilute gas requires knowledge of collision integrals $\Omega^{(l,s)}$ [5,6]. Because of the difficulty of calculating these collision integrals, their values are usually taken from published tables[7,8,9]. Normally, it has been adopted to present these collision integrals as universal functions of reduced temperature. For simple monatomic system the reduced collision integrals has been reported as a function of reduced temperature $T^* = (T/(\varepsilon/k_B))$, where $T$ is the absolute temperature, $\varepsilon$ is the depth of interatomic interaction potential at the position minimum, and $k_B$ is the Boltzman constant. The reported $\Omega^{(l,s)}$ at $T$'s corresponding to $0.3 < T^* < 100$ are quite good for most practical applications[5].

The collision integrals can be calculated numerically on the basis of the Chapman-Enskog solution of the Boltzaman equation if an accurate interatomic interaction potential energy function is known[8]. The collision integrals have been reported for several Lennard-Jones $LJ(m-n)$ [10-13], Buckingham, Exp-6[14,15], Morse[16,17], and Kihara[18] potential functions. The more elaborate among these models have been used to describe the properties of quasispherical and polyatomic molecules with varying degree of success.

Current interest in transport properties has focused attention on the repulsive wall of the pair potential function. It is commonly known that if only the repulsive wall of the potential function is known accurately one is able to determine accurate collision integrals and subsequently the corresponding transport properties within the experimental accuracy[19,20]. This may be completely true in the case of systems composed of closed-shell atoms and molecules. However, in the case of open-shell systems like alkali metals with weak chemical associations, the degree of hardness must be handled carefully. No special procedure is available for such calculation and usually comes as the method imposed by other hard-core systems[21]. The viscosity of liquid metals was investigated in a wide temperature range using the model potential of interatomic interaction within the Enskog-Dymond transport theory developed for a dense hard sphere[22].

Although the current state of art for intermolecular interaction potential allows accurate calculation of the potential energy of the system of interest, the theory of the pairwise interaction, particularly at low density, provides a good estimate of the total interaction analytically by using a pair potential model. The general form of $LJ(m-n)$ potential $u(r)$ is

$$u(r) = \frac{\varepsilon m}{m-n} \left(\frac{m}{n}\right)^{\frac{n}{(m-n)}} \left[\left(\frac{\sigma}{r}\right)^m - \left(\frac{\sigma}{r}\right)^n\right] \tag{1}$$

where $\sigma$ is the hard atomic diameter.

For a large number of fluids that their major interactions between the fluid particles are due to the dispersion, the LJ(12-6) potential function accounts for the pairwise interaction reasonably accurate. The interest in LJ potential is due to its simple analytical form but, particles of many fluids interact with a columbic feature such that the long range dispersion interaction potential is underestimated if treated only by the simple inverse sixth power of the interatomic distances. Although in recent years it has been proved that the three-body interaction is important particularly in condensed phase of argon[23], but it has been recognized that pairwise theory is successful because the three-body repulsion is largely balanced by the corresponding attraction.

The interatomic interaction in alkali metal atoms changes from a screened columbic potential in solid state to the LJ-type interaction in vapor state [24,25]. In a series of investigations, it has been shown that the LJ(6-3) potential model, can be applied to predict the thermodynamic properties of liquid cesium metal quite accurately[26]. The form of the repulsion wall has been estimated by the application of neutron scattering data of liquid cesium metal[27,28], and the attraction branch has been inspired by the form of dipole-dipole interaction potential[29,30]. In the same way, Exp-6 potential function has been shown to describe the properties of cesium liquid metal quite accurately[31]. Also LJ(12-6) potential function has been used to predict the thermodynamic properties of liquid alkali metals at any temperature above a given liquid density, for instance 1.21 g/cm$^3$ in the case of liquid cesium metal, below which substantial deviations are seen[32]. In 1991, Kozhevnikov et al. have used their experimental *PVT* data of liquid cesium metal in the range 400-2000 K and accordingly have reported the LJ(8.5-4) potential function through the calculation of the cohesive energy density[33]. Thus the repulsive wall of these potentials involves a softer repulsion in short range then a hard-sphere like argon. The application of LJ(8.5-4) potential function has been shown to predict transport properties more accurately than LJ(6-3) potential, and thus LJ(8.5-4) was superior potential for having a qualified harder repulsion[34]. Although the LJ(8.5-4) potential function has been able to predict the equilibrium thermodynamic properties quite accurately, it has been failed an accurate prediction of transport properties such as viscosity and thermal conductivity of cesium vapor. This can be attributed to the improper steepness of the repulsion wall of the LJ(8.5-4) potential function.

It is well-known that transport properties, especially viscosity, can be determined principally by the repulsion wall of the potential function. In particular, it has been demonstrated that the hard-sphere model with the packing density $\eta_m$= 0.48 at the melting temperature reproduces well the viscosity of (almost all) melts in which the metallic bond predominates [22]. From this view, in this work, initially we have used LJ(12-6) potential function for prediction of transport properties of K, Rb, and Cs metals vapor but it failed having good agreement with experiment especially in the low temperature range.

In this study to promote the accurate prediction of viscosity and thermal conductivity in a wide range of temperature, we use the potential function LJ(15-6), which is harder than LJ(12-6). Knowing that this potential is not suitable for the description of the equilibrium thermodynamics of the metallic system, we determine the required molecular potential parameters in a separate study on the equation of state of alkalis. The reduced collision integrals are calculated using

LJ(15-6) potential function. To study an additional degree of flexibility for the description of the core-core interaction in the calculation of the transport properties in a wider temperature range forms another purpose of the present paper.

## 2. The method

The values of reduced collision integrals $\Omega^{(2,2)*}$ and $\Omega^{(2,3)*}$ are calculated from Chapman-Enskog solution of the Boltzman equation by using the LJ(15-6) potential function. For the calculation of the collision integrals, the algorithm of O'Hara and Smith is used[35].

The molecular potential parameters $\varepsilon$ and $\sigma$, which are required in the calculations of transport properties, depend on the chosen potential model. It has been already established that the available compressed *PVT* data of liquid cesium metal from freezing point to the critical point can be modeled by LJ(8.5-4) potential function quite accurately. Using this model potential, an analytical equation of state (EOS) is derived and then the isotherm of the equation of state, which is perfectly linear, is used to calculate the potential parameters over the temperature range for which *PVT* data is available. Hence the compressed liquid *PVT* data of the alkali metals are the only input data[36,37]. The readers are referred to references 26 and 31 for the details. The fact that LJ(8.5-4) potential is capable of predicting the equilibrium thermodynamic properties quite accurately but not the transport properties rationalizes the simulation of a hard-wall potential function from available information which at the same time grants access to the thermodynamic properties. Using the experimental viscosities of K, Rb, and Cs and their related $\varepsilon$ and $\sigma$ of LJ(8.5-4) potential function, the values of reduced collision integrals as functions of $T^*$ for the three metals are calculated. Then, using the algorithm of O'Hara and Smith, $\Omega^{(2,2)*}$'s and $\Omega^{(2,3)*}$'s are calculated for LJ(6-3), LJ(8.5-4) and LJ(15-6) potential functions. The same procedure is followed by using experimental thermal conductivities. Then inquiring about the law of corresponding states, LJ(15-6) potential function finds the best match of both viscosity and thermal conductivity. With this rational and the method outlined above, a kind of *hybrid* potential with steep short-range wall suitable for the calculation of the transport properties is obtained for that metal. This LJ(15-6) hybrid potential function (hereafter hybrid potential) maintains mainly the long range dispersion interaction due to application of the experimental *PVT* data and the short range interaction due to the choice of hard wall character of LJ(15-6) potential.

## 3. Numerical Calculation

### 3-1. Collision integral: Law of Corresponding States

For dilute gases, the transport coefficients essentially depend on the binary collision interaction between molecules. The Chapman-Enskog solution of the Boltzman kinetic equation for dilute monatomic gases relates these coefficients to a series of reduce collision integrals $\Omega^{(l,s)*}$. In the kinetic theory expressions for the dilute gas transport properties, the interaction potential function $u(r)$, is covered under three layers of integration, which makes it appears that the values of transport coefficients should reflect the details of $u(r)$ precisely.

Along with studying the calculation of transport coefficients, the law of corresponding states as a universal function can be expected to hold. According to the theory of the equation of state, for fluid system whose interaction potential can be expressed by a two-parameter function, e.g., $u \propto \varepsilon f(\sigma/r)$, there exist a universal function for the property of interest as a function of reduced temperature $T^*(= T/(\varepsilon/k_B))$ and the reduced distance $r^*(= r/\sigma)$. Since this, then it can be proposed that for the LJ(15-6) potential $\Omega^{(l,s)*} \propto f(T^*)$, where $\Omega^{(l,s)*} = \Omega^{(l,s)}/\pi\sigma^2$. The calculated values of $\Omega^{(l,s)*}$ as a function of $T^*$ can be used to seek the universality of alkali metals. The correlation of reduced collision integrals versus $T^*$ of various metals based on experimental-transport, follow the trend of a same universal function, which is nicely approximated by the correlation based on LJ(15-6) potential. In this study an accurate correlation between reduced collision integrals and $T^*$ is established, which qualifies the LJ(15-6) potential for presentation of hybrid potential and the calculation of accurate viscosity and thermal conductivity over a wide range of temperature.

**4- Result and Discussion**

**4-1 Potential Function and Molecular Parameters**

The transport properties of a fluid are due to interparticle collisions, and therefore they must depend on the details of interaction potential at the short range. Most theoretical investigations for thermophysical properties have shown that transport properties are sensitive to the steepness of the repulsion wall of the potential function [22]. On the other hand the equilibrium thermodynamic properties are sensitive to the asymptotic form of the potential function at long distance. As a practical experience a potential function, which is useful for the accurate prediction of density, for instance, could not necessarily predict the viscosity with a great accuracy. From this view, therefore LJ(8.5-4) needs to be more justified for application in the calculation of transports.

First the intermolecular potential parameters $\varepsilon$ and $\sigma$ are determined from experimental *PVT* data of compressed liquid metal in the range 500-1600 K. These are obtained by using the slope and intercept of the linear isotherms based on the LJ(8.5-4) potential function. Since the LJ(8.5-4) potential function produces perfect linear isotherms over the whole liquid range,[34] it can be concluded that it is the best fit to the equilibrium (experimental *PVT*) data. On the contrary, the isotherm of the equation of state based on the harder LJ(15-6) potential function does not show linear behavior even with moderate accuracies. This can be attributed to the fact that the equilibrium properties mainly correspond to the average intermolecular distances around the corresponding minimum position of the potential. For instance the liquid volumes (of the available *PVT* data) correspond to the intermolecular distances between $0.9r_m$ to $1.4r_m$, where $r_m$ is the position of the potential minimum. Therefore these data are belong to the mid- and long-ranges, which is described suitably by an attractive potential like LJ(8.5-4), and do not cover much of the short range. Equivalently, fitting in the available *PVT* data, the molecular potential parameters of hard-wall LJ(15-6) potential model neither produce equilibrium nor transport properties of the metals with reasonable accuracies.

Next, for each fluid metal, a hybrid potential is produced by using the corresponding average values of $\varepsilon$ and $\sigma$ of LJ(8.5-4) potential (obtained above) in the LJ(15-6) potential function. This is easily justified by the fact that $\varepsilon$ and particularly $\sigma$ depend slightly on the temperature. The average values of $\varepsilon/k_B$ and $\sigma$ for K, Rb, and Cs metals are shown in Table 1.

The hybrid potential is shown in Figure 1 typically for cesium metal, and compared with original LJ(15-6) and LJ(8.5-4) potentials. From the Figure 1, it can be seen that the hybrid potential is almost as steep as the original LJ(15-6) potential function, but it is as penetrable as LJ(8.5-4). Since the $ns^1$ electron clouds of the two colliding alkali metal atoms are to a good deal penetrable when they encounter a collision, then it can be seen that the hybrid potential describes the interatomic interaction of the alkali metals quite well. This is the rational for using LJ(8.5-4) for simulating the hybrid potential, which transfers the size-parameter effectively.

**Table 1.** Average molecular parameters calculated by using LJ(8.5-4) potential function for K, Rb, and Cs metals.

| Parameters | K | Rb | Cs |
|---|---|---|---|
| $\sigma(\text{Å})$ | 4.137 | 4.405 | 4.764 |
| $\varepsilon/k_B(\text{K})$ | 974.89 | 901.57 | 842.157 |

Notice that the position of the minimum in hybrid potential is shifted to a lower distance as compared to both the original LJ(15-6) and LJ(8.5-4). Furthermore, the hybrid potential has a smaller attraction than LJ(8.5-4) at the long range. Here the potential parameters $\sigma$ and $\varepsilon$ are acting as constraints and transfer the character of the liquid cesium metal. In particular, the constraint $\sigma$ not only fixes the size of atom-atom interaction at contact, but also it is responsible for shifting the position of potential minimum $r_m$ to a lower distance as demanded and regulated by a harder repulsion wall of LJ(15-6) compared to LJ(8.5-4). It is interesting to note that, for cesium as an example, the position of minimum of the hybrid potential ($r_m = 5.30 \dot{A}$) is in excellent agreement with the result of neutron scattering of the liquid state near the freezing point ($r_m = 5.31 \dot{A}$) [38].

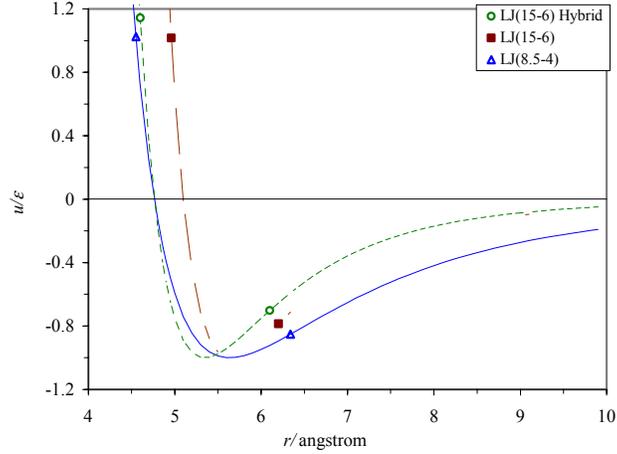

**Figure 1.** The potential functions for cesium metal (see text for details).

Certainly a potential model characterized by using the experimental data would produce parameters ($\varepsilon$ and $\sigma$) characteristics of the model.

**4-2. Collision Integral**

To the first approximation, the collision integral $\Omega^{(2,2)}$ corresponds to viscosity and thermal conductivity. Higher approximations are obtained by multiplying the first approximation for the transport coefficients by particular correction factors, which are near unity. The viscosity correction factor $f_\eta^{(2)}$, and thermal conductivity correction factor $f_\lambda^{(2)}$, which are calculated by using values of $\Omega^{(2,2)*}$ and $\Omega^{(2,3)*}$, are commonly used[39].

It is valuable to present the collision integrals as a universal function in the context of the law of corresponding states for validation and prediction. By the help of calculated potential parameters ($\varepsilon$ and $\sigma$) based on LJ(8.5-4) potential, the experimental viscosity and thermal conductivity can be used to determine the reduced collision integrals as a function of reduced temperature. Here, we have used viscosities and thermal conductivities data of reference 40. Using these values, we obtain $\Omega_\eta^*$ and $\Omega_\lambda^*$, where $\Omega_\eta^*(=\Omega^{(2,2)*}/f_\eta^{(2)})$ and $\Omega_\lambda^*(=\Omega^{(2,2)*}/f_\lambda^{(2)})$ (see Eqs. 2 and 3 in the next section). The correlation of $\Omega_\eta^*$ and $\Omega_\lambda^*$ as functions of $T^*$ for various metals are shown in Figure 2 and 3, respectively.

On the other hand, using the values of $\Omega^{(2,2)*}$ and $\Omega^{(2,3)*}$ based on the LJ(15-6) potential, we obtain correlations of $\Omega_\eta^*$ and $\Omega_\lambda^*$ versus $T^*$. The relation of $\Omega_\eta^*$ and $\Omega_\lambda^*$ are smooth functions of $T^*$ suitable for accurate numerical interpolation and extrapolation. The results are shown in Figures 2 and 3, respectively. Also, based on LJ(8.5-4) and LJ(6-3) potentials, the corresponding correlations of $\Omega_\eta^*$'s and $\Omega_\lambda^*$'s are calculated and depicted in the same Figures for comparison. It can be seen that the correlations of $\Omega_\eta^*$'s as functions of $T^*$, based on experimental viscosity and the application of $\varepsilon$ and $\sigma$ of the LJ(8.5-4) potential function for the various alkalis, follow the same trend and exhibit a universal behavior. Inspection of Figure 3 (and the insert), again shows that only the $\Omega_\lambda^*$ based on the LJ(15-6) potential predicts the universality shown by $\Omega_\lambda^*$'s of the various metals based on the experiment thermal conductivities, though with less accuracy than with viscosities. Slight deviations are seen in Figures 2 and 3, which can be attributed to the difference in the electronic structure of the alkali metals as one move down the periodic table[22].

The correlation between three alkali metals in Figure 3 is not strong and can be attributed to relatively high inaccuracies in the experimental thermal conductivity as compared to the experimental viscosity. The important conclusion comes by the inspection of Figures 2 and 3, and noting that the resulted $\Omega_\eta^*$ and $\Omega_\lambda^*$ based on the LJ(6-3) and LJ(8.5-4) potentials, are not in agreement with the experiments even with moderate accuracies. Therefore, by adjusting the steepness of the repulsion wall, the transport properties can be calculated with high accuracies. This conclusion may be used to validate the previous finding of extensive dependence of the transports to the repulsion wall of the pair interaction potential function[8,22].

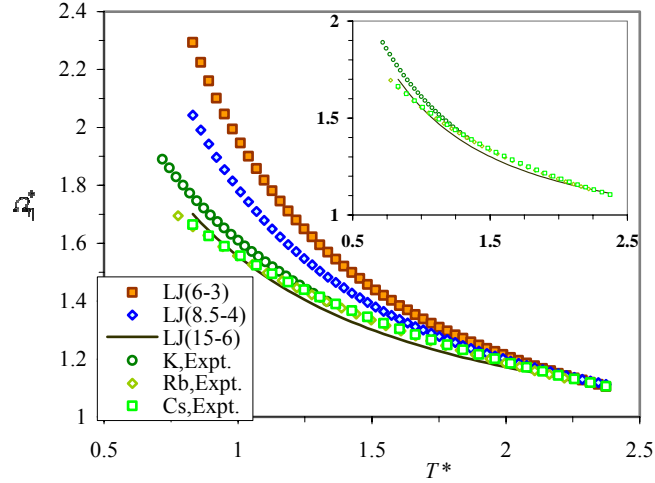

**Figure 2.** Plots of $\Omega_\eta^*$'s based on experimental viscosity and LJ(*m-n*) potential for K, Rb, and Cs metals.

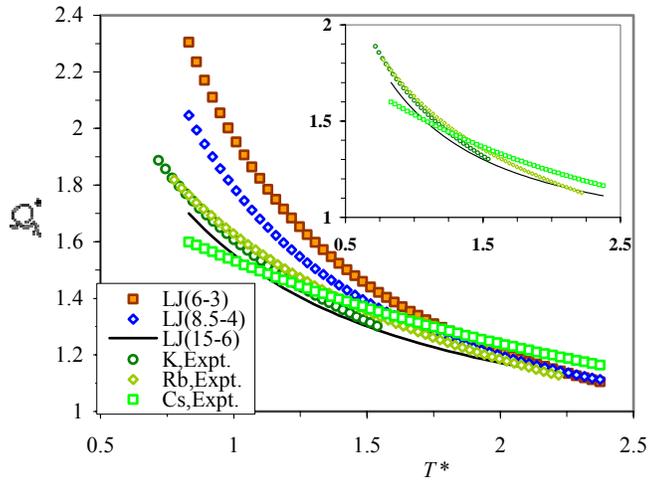

**Figure 3**. Plots of $\Omega_\lambda^*$'s based on experimental thermal conductivity and LJ(*m-n*) potential for K, Rb, and Cs metals.

In short, $\Omega_\eta^*$'s and $\Omega_\lambda^*$'s based on experimental transports of K, Rb, and Cs metals vapor demonstrate the law of corresponding states, and follow a universal function that can be nicely predicted by the correlation of collision integrals based on the LJ(15-6) potential. Understandably, the hybrid potential is in the sense that the $\Omega^{(2,2)*}$ based on the LJ(15-6) potential is the central quantity for the calculation of transport properties using $\varepsilon$ and $\sigma$ of LJ(8.5-4) potential, calculated and shown in Table 1.

**4-3 Viscosity and Thermal conductivity**

The interpretation of the experimental transport data of the alkali metal vapors is complicated because of *n*-mers formation even at low pressures. Therefore, the theoretical calculations of transport properties of alkali metals based on the monatomic species should provide valuable information only on the zero density limits of the experimental transport data. In this study, we calculate the transport properties for a particular fluid system consisting of monatomic particles.

According to the kinetic theory, the viscosity of a gas at the limit of zero density $\eta(T)$, can be predicted by[10,41].

$$\eta(T) = \frac{5}{16}\left(\frac{mk_BT}{\pi}\right)^{1/2}\frac{1}{\sigma^2\Omega_\eta^*}, \qquad \Omega_\eta^* = \Omega^{(2,2)*}/f_\eta^{(2)} \qquad (2)$$

where $m$ is the molecular mass. The transport collision integrals $\Omega^{(2,2)*}$ and $\Omega^{(2,3)*}$ (required for the calculation of $f_\eta^{(2)}$ and $f_\lambda^{(2)}$) are calculated for the LJ(15-6) potential, using the algorithm given by O'Hara and Smith [35]. Notice that here the values of $\sigma$ are taken from Table 1.

In the same way, the thermal conductivity $\lambda(T)$, can be calculated by[10,41]

$$\lambda(T) = \frac{25}{32}\left(\frac{mk_BT}{\pi}\right)^{1/2}\frac{C_v}{\sigma^2\Omega_\lambda^*}, \qquad \Omega_\lambda^* = \Omega^{(2,2)*}/f_\lambda^{(2)} \qquad (3)$$

where $C_v$ is the heat capacity. The heat capacity is contributed by monomer and dimer as well as higher-mers. However, it has been known that the fraction of monomers is considerably larger than the dimers. Therefore, according the above approximation, the heat capacity is contributed only by the translational energy of the alkali metal monomer.

**Table 2**. Calculated %AAD of viscosity and thermal conductivity for K, Rb, and Cs metals.

| Temperature Range (K) | %AAD $\eta$ | %AAD $\lambda$ |
|---|---|---|
| Potassium | | |
| 700-1500 | 2.9704 | 2.7828 |
| Rubidium | | |
| 700-2000 | 1.6937 | 3.2592 |
| Cesium | | |
| 700-2000 | 1.7538 | 3.6294 |

In Table 2, values of percent Average Absolute Deviation (%AAD) of viscosities and thermal conductivities for K, Rb, and Cs metal vapors are presented. Deviations are with respect to the experimental viscosity and thermal conductivity data of monomers reported by Vargaftik [40]. Prediction of viscosity is in good agreement with experimental and the maximum %AAD (=2.97) is occurred in the case of potassium. Also, the thermal conductivity values are in good agreement with the experiment and the maximum %AAD(=3.62) is due to cesium.

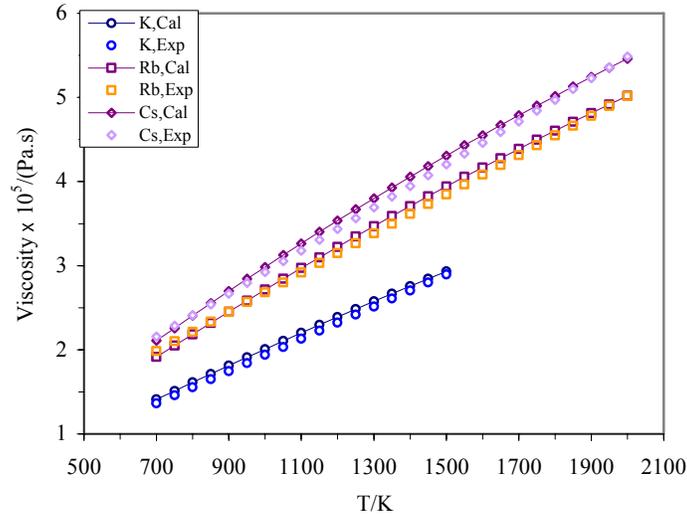

**Figure 4**. Comparison of the predicted viscosity with the experimental viscosity data of K, Rb, and Cs metals vapor.

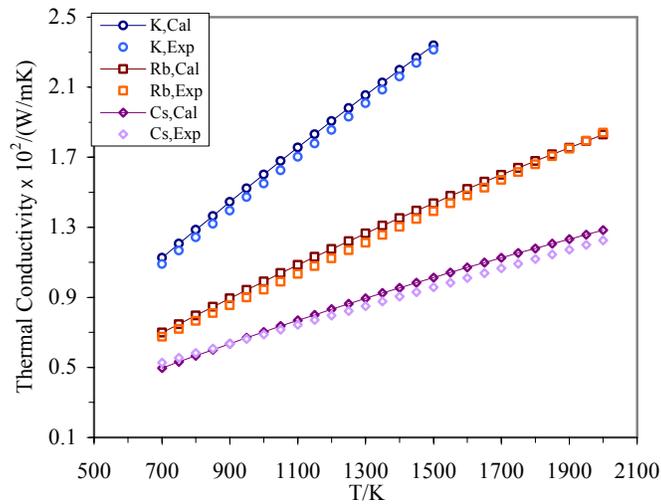

**Figure 5**. Comparison of predicted thermal conductivity with experimental thermal conductivity data of K, Rb, and Cs metals vapor.

It should be mentioned that the thermal conductivity measurement is generally subjected to a higher error than the viscosity measurement. In Figure 4, the values of viscosity based on the hybrid potential are compared with the available experimental values. In general, the maximum deviation in viscosity occurs in the mid temperature range for the three metals. More or less this is the case for the predicted thermal conductivity, which are compared with experiment and shown in Figure 5.

**5- Conclusion**

The two-parameter hard-wall hybrid potential function has been used to predict viscosity and thermal conductivity of K, Rb, and Cs metal vapors. The experimental-transport-based $\Omega_\eta^*$ and $\Omega_\lambda^*$ as a function of reduced temperature of the three alkalis demonstrate that the law of corresponding states, and follow a same universal correlation that is predicted accurately by the hard-wall hybrid potential. This rationalizes the applicability of hard-wall hybrid potential for describing the transport properties of alkali vapors. Values of molecular parameters of the hybrid potential, $\sigma$ and $\varepsilon$, which have been calculated from *PVT* data of the liquid state by using the LJ(8.5-4) potential function, suitably presents a method for the calculation of alkalis transport.

It has been shown that hybrid potential can be applied successfully to predict the viscosity and thermal conductivity of K, Rb, and Cs vapor in the temperature range 700-2000 K accurately, in particular in low temperature range. The hybrid potential is involved a repulsive wall harder than both LJ(8.5-4) and LJ(6-3) potential functions, and is able to predict transport properties with good accuracy. The results give the impression that transport properties of potassium, rubidium, and cesium vapors are highly sensitive to the repulsive wall of the potential function.

The relation of $\Omega_\eta^*$ and $\Omega_\lambda^*$ based on the LJ(15-6) potential are smooth functions of $T^*$ suitable for accurate numerical interpolation and extrapolation. They can be used for calculation of viscosity and thermal conductivity without involving the calculation of $f_\eta^{(2)}$ and $f_\lambda^{(2)}$ correction factors.

**Acknowledgment**

The authors are indebted to the Research Councils of Shiraz University for supporting this project.**References**

1. N.B. Vargaftik, and V.S. Yargin, in: R.W. Ohse (Ed.), Handbook of Thermodynamic and Transport Properties of Alkali Metals, Academic Press, Oxford, P. 485 (1985).13

Figure Captions

**Figure 1.** The potential functions for cesium (see text for details).

**Figure 2.** Plots of $\Omega_\eta^*$'s based on experimental viscosity and LJ(*m-n*) potential for K, Rb, and Cs metals.

**Figure 3**. Plots of $\Omega_\lambda^*$'s based on experimental thermal conductivity and LJ(*m-n*) potential for K, Rb, and Cs metals.

**Figure 4**. Comparison of the predicted viscosity with the experimental viscosity data of K, Rb, and Cs metals vapor.

**Figure 5**. Comparison of predicted thermal conductivity with experimental thermal conductivity data of K, Rb, and Cs metals vapor.

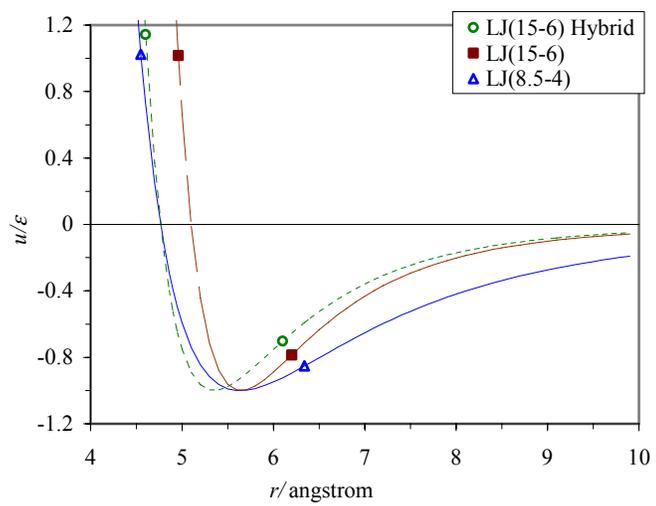

Ghatee and Niroomand                     Figure 1

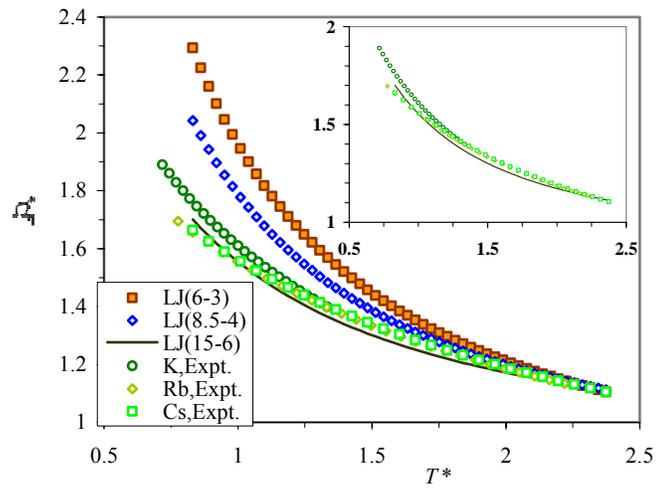

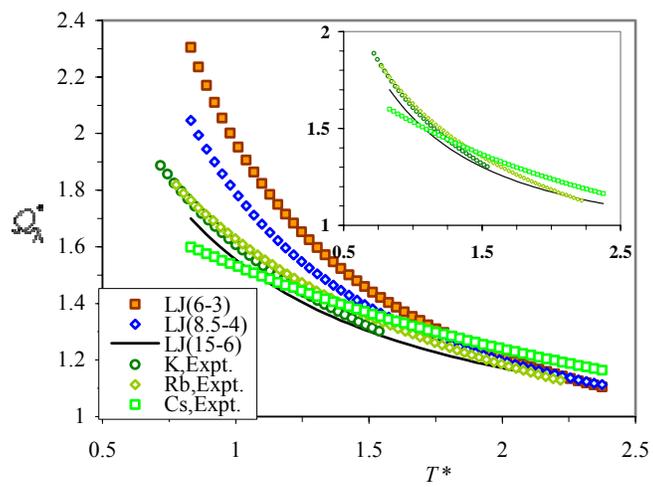

Ghatee and Niroomand          Figure 3

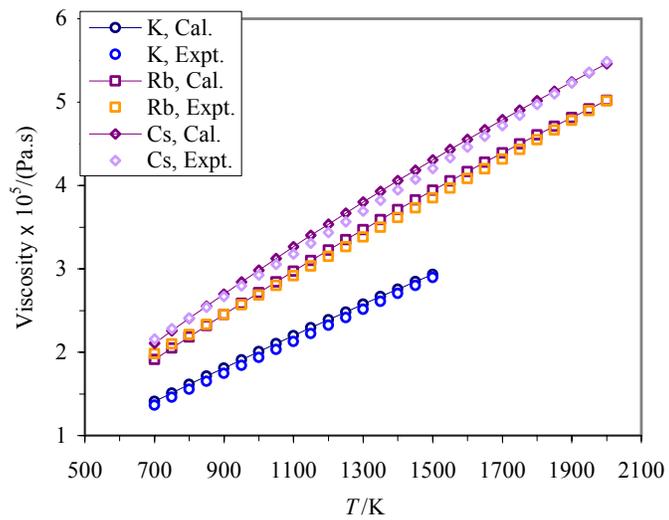

Ghatee and Niroomand                    Figure 4

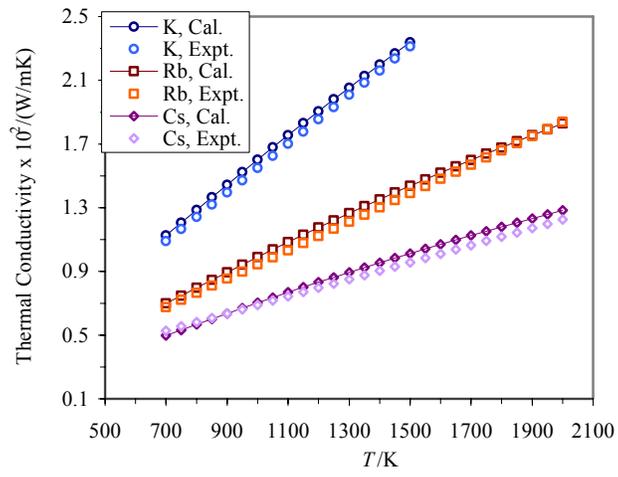

Ghatee and Niroomand                    Figure 5